\documentclass[submission,copyright,creativecommons]{eptcs}
\providecommand{\event}{EXPRESS/SOS 2019} % Name of the event you are submitting to
\usepackage{amssymb}
\usepackage{amsthm}
\usepackage{stmaryrd}
\usepackage{tikz}
\usetikzlibrary{trees,arrows,arrows.meta,shapes,decorations.pathmorphing,backgrounds,positioning,fit,petri}
\usepackage{algorithm}
\usepackage[noend]{algpseudocode}
\usepackage{subcaption}
\theoremstyle{plain}

\newtheorem{theorem}{Theorem}[section]
\newtheorem*{theorem*}{Theorem}
\newtheorem{lemma}[theorem]{Lemma}
\newtheorem*{lemma*}{Lemma}
\newtheorem{proposition}[theorem]{Proposition}

\newtheorem{definition}[theorem]{Definition}
\theoremstyle{definition}
\newtheorem{example}[theorem]{Example}
\newcommand{\be}{\begin{example}}
\newcommand{\ee} {\qed \end{example}}
\theoremstyle{remark}
\newtheorem{remark}{Remark}[section]

\newcommand{\Kop}[2]{\mathsf{K}_{#1}(#2)}
\newcommand{\Gop}[2]{\mathsf{D}_{#1}(#2)}

\newcommand{\sfunspace}[1]{\mathcal{S}(#1)}  % Set of all space functions
\newcommand{\sfuntuple}{\mathfrak{s}}  
\newcommand{\efuntuple}{\mathfrak{e}}
\newcommand{\pfuntuple}{\pi}
\newcommand{\Pfuntuple}{\Pi}
\newcommand{\Dfuntuple}{\Delta}

\newcommand{\sfun}[1]{\sfuntuple_{#1}}
\newcommand{\efun}[1]{\efuntuple_{#1}}	
\newcommand{\pfun}[1]{\pfuntuple_{#1}}	
\newcommand{\Pfun}[1]{\Pfuntuple_{#1}}	
\newcommand{\Dfun}[1]{\Dfuntuple_{#1}}

\newcommand{\sfunapp}[2]{\sfun{#1}( {#2} )}		
\newcommand{\efunapp}[2]{\efun{#1}( {#2} )}
\newcommand{\pfunapp}[2]{\pfun{#1}( {#2} )}
\newcommand{\Pfunapp}[2]{\Pfun{#1}( {#2} )}
\newcommand{\Dfunapp}[2]{\Dfun{#1}( {#2} )}

\newcommand{\C}{\mathbf{C}}         	 % Constraint System Variable
\newcommand{\Con}{\mbox{\it Con}}   	 % Constraint set
\newcommand{\true}{\it{true}}			 % True
\newcommand{\false}{\it{false}}			 % False
\newcommand{\join}{\sqcup}	 			 % Join operator
\newcommand{\bigjoin}{\bigsqcup}		 % Join operator
\newcommand{\meet}{\sqcap}	 			 % Meet operator
\newcommand{\bigmeet}{\bigsqcap}		 % Meet operator
\newcommand{\cleq}{\sqsubseteq} 		 % Order relation
\newcommand{\cgeq}{\sqsupseteq} 		 % Inv. order relation
\newcommand{\cg}{\sqsupset} 			 % strict Inv. order relation
\newcommand{\Cs}{\C_{{\texttt s}}}  	 % Space Function Lattice Variable
\newcommand{\fleq}{\cleq_{{\texttt s}}}  % Function order
\newcommand{\imp}{\rightarrow}

\newcommand{\defsymbol}{\stackrel{\textup{\texttt{def}}}  {=}}

\newcommand{\acal}{\mathcal{A}}
\newcommand{\pcal}{\mathcal{P}}

\newcommand{\nmat}{\mathbb{N}}

\providecommand{\doi}[1]{\textsc{doi}: \href{http://dx.doi.org/#1}{\nolinkurl{#1}}}
\usepackage{hyperref}

\title{Semantic Structures for Spatially-Distributed Multi-Agent Systems}
\author{Frank Valencia\thanks{Frank Valencia has been partially supported by the ECOS-NORD project FACTS (C19M03).}
\institute{CNRS-LIX Ecole Polytechnique de Paris}
\institute{Univ. Javeriana Cali}
\email{frank.valencia@lix.polytechnique.fr}
}
\def\titlerunning{Semantic Structures for Spatially-Distributed Multi-Agent Systems}
\def\authorrunning{F. Valencia}
\begin{document}
\maketitle

\begin{abstract}
Spatial constraint systems (scs) are semantic structures for reasoning about
spatial and epistemic information in concurrent systems. They have been
used to reason about beliefs, lies, and group epistemic behaviour inspired by social networks. They have also been
used for proving new results about modal logics and giving semantics to process calculi. 
In this paper we will discuss the theory and main results about scs.
\end{abstract}

\section{Introduction}
\label{sec:intro}

Epistemic, mobile and spatial behavior are common place in today's distributed systems.  The intrinsic \emph{epistemic} nature of these systems arises from social behavior. Most people are familiar with digital systems where \emph{agents} (users) share their \emph{beliefs}, \emph{opinions}  and even intentional \emph{lies} (hoaxes).  Also, systems modeling decision behavior must account for those decisions' dependance on the results of interactions with others within some social context. The courses of action stemming from some agent decision result  not only from the rational analysis of a particular situation but also from the agent beliefs  or information that sprang from the interactions with other participants involved in that situation. %Besides being epistemic these systems are also \emph{higher-order}: 
Appropriate performance within these social contexts requires the agent to form beliefs about the beliefs of others.  %on social networks.   
Spatial and mobile behavior is exhibited by apps and data moving across (possibly nested) spaces defined by, for example,  friend circles 
 and shared folders. We therefore believe that a solid understanding of the notion of \emph{space} and \emph{spatial mobility} as well as the flow of epistemic information is relevant in any model of today's distributed systems.
 
The notion of group is also fundamental in distributed systems. Since the early days of multi-user operating systems,  information was categorized into
that available to one user, some group of users, or everyone. Information was thus
separated into ``spaces'' with  boundaries defined by accessibility. In these systems
we could say that, from the restrictive point of view of  information ``permissions'',
the notion of group was \emph{reified} as another agent of the system.

In current distributed systems such as social networks, actors behave more as
members of a certain \emph{group} than as isolated individuals. Information,
opinions, and beliefs of a particular actor are frequently the result of an
evolving process of interchanges with other actors in a group. This suggests a reified notion of group as a single actor operating within the
context of the collective information of its members. It also conveys two notions of information, one spatial and the other epistemic. In the former, information is localized in compartments associated with a
user or group. In the latter, it refers to something known or believed by a
single agent or collectively by a group.
%
%In a previous CONCUR paper~\cite{knight2012spatial} the authors proposed
%\emph{spatial constraint systems (scs)}, an extension of the semantic structures for
%\emph{concurrent constraint programming (ccp)} paradigm~\cite{saraswat1991semantic},
%to account for both, spatial and epistemic information, in which agents are self-maps
%in the lattice of information. One of the goals of that work, following a plea made
%by Panangaden~\cite{Panangaden08} at a joint PODC-CONCUR conference in 2008,
%was to use epistemic principles in concurrency theory. 

Furthermore, in many real life multi-agent systems, the  agents are unknown in
advance.  New agents can subscribe to the system in
unpredictable ways.  Thus, there is usually no a-priori bound on the number of agents in the  system. It is then often convenient to model the group of agents as an infinite set. 
In fact, in models from economics and epistemic logic \cite{Hildenbrand1970,halpern2004reasoning}, groups of agents have been represented as infinite, even uncountable, sets.
This raises interesting issues about the distributed  information of such groups. In particular, that
of \emph{group compactness}: information that when obtained by an infinite
group can also be obtained by one of its finite subgroups. 

Spatial constraint systems (scs)\footnote{For simplicity we use
\emph{scs} for both \emph{spatial constraint system} and its plural form.} are semantic structures for the epistemic behaviour of multi-agent systems. 
These structures single out the notions we previously discussed: Namely, space, beliefs, and distributed information of potentially infinite groups.  In this paper we will describe  the theory of scs and highlight its main results from \cite{guzman:hal-01257113,guzman:hal-01328188,guzman:hal-01675010,haar:hal-01256984,guzman:hal-02172415}. 
 
 \section{Overview}
 
 In this section we will give a brief description and motivate scs in the context of space, extrusion, and distributed information.

{Declarative} formalisms of concurrency theory such as process calculi for {\it concurrent constraint programming} (ccp)~\cite{saraswat1991semantic} were designed to give explicit 
access to the concept of {partial information} and, as such, have close ties with logic.  
This makes them ideal for the incorporation of epistemic  and spatial concepts by expanding the logical connections to include \emph{multi-agent modal logic}~\cite{kripke1963semantical}.  
 In fact, the  sccp calculus~\cite{knight:hal-00761116} extends ccp with the ability to define local computational spaces where agents can store epistemic information and run processes. 

Constraint systems (cs) are algebraic structures  for the semantics of  ccp  \cite{saraswat1991semantic,de1995nondeterminism,knight:hal-00761116,Fages01,Rety98,nielsen2002temporal}.  
They specify the domain and elementary operations and partial information upon which programs (processes) of these
calculi may act. 

A  cs can be formalized as a complete lattice $(\Con, \cleq)$. The elements of
$\Con$ represent partial  information and we shall think of them as being
\emph{assertions}.  They are traditionally referred to as \emph{constraints}
since they naturally express partial information (e.g., $x>42$). The order
$\cleq$ corresponds to entailment between constraints, $c \cleq d$, often
written $d \cgeq c$, means $c$ can be derived from $d$, or that $d$ represents
as much information as $c$. The join $\join$, the bottom $\true$, and the top
$\false$ of the lattice correspond to conjunction, the empty information, and
the join of all (possibly inconsistent)  information, respectively.  

Constraint systems provide the domains and operations upon which the semantic foundations of ccp calculi are built. As such,  ccp operations and their logical counterparts typically  have a corresponding 
elementary construct or operation on the elements of the constraint system. In particular, parallel composition and conjunction correspond  to the \emph{join} operation, and existential quantification and 
local variables correspond to a cylindrification operation on the set of constraints~\cite{saraswat1991semantic}. 

{\it Space.} Similarly, the notion of computational space and the epistemic notion of belief in sccp~\cite{knight:hal-00761116}  correspond to a family of join-preserving maps $\sfun{i}:\Con \rightarrow \Con $ called
\emph{space functions}.  A cs equipped with space functions is called a
\emph{spatial constraint system} (scs). From a \emph{computational point of
view}  $\sfunapp{i}{c}$ can be interpreted as an assertion specifying that 
$c$ resides within the space of agent $i.$ From an \emph{epistemic point of
view}, $\sfunapp{i}{c}$ specifies that  $i$ considers $c$ to be true. An
alternative epistemic view is that $i$ interprets $c$ as $\sfunapp{i}{c}$. All
these interpretations convey the idea of $c$ being local or subjective to
agent $i$.
%We have generalized cs to reason about epistemic, mobile and spatial behavior \cite{guzman:hal-01257113,guzman:hal-01328188,guzman:hal-01675010,haar:hal-01256984,guzman:hal-02172415} 

 In the spatial ccp process calculus
\emph{sccp}~\cite{knight:hal-00761116},  scs are used to specify the spatial
distribution of information in configurations $\langle P, c \rangle$ where $P$
is a process and $c$ is a constraint, called \emph{the store},  representing
the current partial information. E.g.,  a reduction \( \langle \ P,
{\sfunapp{1}{a}}\join {\sfunapp{2}{b}} \ \rangle \longrightarrow \langle\ Q,
{\sfunapp{1}{a}} \join {\sfunapp{2}{b \join c}} \ \rangle \) means that $P$,
with $a$ in the space of agent $1$ and $b$ in the space of agent $2$, can
evolve to $Q$ while adding $c$ to the space of agent $2$.

{\it Extrusion.}  An extrusion function for the space $\sfun{i}$ is a map $\efun{i}:\Con \rightarrow \Con$ that satisfies \( \ \sfunapp{i}{\efunapp{i}{c}} = c\).  This means that we think of extrusion  as the \emph{right inverse} of space.  Intuitively, within a space context $\sfunapp{i}{\cdot}$, the assertion $\efunapp{i}{c}$ specifies that $c$ must be posted outside  of agent $i$'s space.  
 The computational interpretation of $\efun{i}$ is that of a process being able to extrude any $c$ from the space  $\sfun{i}.$ 
The extruded information $c$ may not necessarily be part of the information residing in the space of agent $i$. For example, using properties of space and extrusion
functions we shall see that $\sfunapp{i}{\ d \sqcup \efunapp{i}{c}} = \sfunapp{i}{d}  \sqcup c$ specifying that $c$ is extruded (while $d$ is still in the space of $i$).  The extruded $c$  could be inconsistent with $d$ (i.e., $c \sqcup d =\false$), it could be related to $d$ (e.g., $c  \sqsubseteq d$), or simply unrelated to $d$. From an epistemic perspective, we can use $\efun{i}$  to express  \emph{utterances} by agent $i$ and such utterances could be intentional lies (i.e., inconsistent with their beliefs), informed opinions (i.e., derived from the beliefs), or simply arbitrary statements (i.e., unrelated to their beliefs). 

{\it Distributed Information.} Let us consider again the sccp reduction \( \langle \ P,
{\sfunapp{1}{a}}\join {\sfunapp{2}{b}} \ \rangle \longrightarrow \langle\ Q,
{\sfunapp{1}{a}} \join {\sfunapp{2}{b \join c}} \ \rangle \). Assume that $d$ is some piece of information
resulting from the combination (join) of the three constraints above,  i.e.,
$d  = a \join b \join c$,  but strictly above the join of any two of them. We
are then in the situation where neither agent has $d$ in their spaces, but as
a group they could potentially have $d$ by combining their information.
Intuitively, $d$ is distributed in the spaces of the group $I= \{ 1, 2 \}$.
Being able to predict the information that agents $1$ and $2$ may derive as
group is a relevant issue in multi-agent concurrent systems,  particularly if
$d$ represents unwanted or conflicting information (e.g., $d=\false$). 

In \cite{guzman:hal-02172415} we introduced the theory of group space functions $\Dfun{I}:\Con
\to \Con$ to reason about information distributed among the members of a
potentially infinite group  $I$. We refer to $\Dfun{I}$ as the
\emph{distributed space} of group $I$. In our theory $c \cgeq \Dfunapp{I}{e}$
holds exactly when we can derive from $c$ that $e$ is distributed among the
agents in $I$.  E.g., for $d$ above, we should have ${\sfunapp{1}{a}} \join
{\sfunapp{2}{b \join c}} \cgeq \Dfunapp{\{1,2\}}{d} $ meaning that from the
information  ${\sfunapp{1}{a}} \join \sfunapp{2}{b \join c}$ we can derive
that $d$ is distributed among the group  $I= \{ 1, 2 \}$.  Furthermore, 
$\Dfunapp{I}{e} \cgeq \Dfunapp{J}{e}$ holds whenever $I \subseteq J$ since if
$e$ is distributed among a group $I$, it should also be distributed in a group
that includes the agents of $I$.

Distributed information of infinite groups can be used to reason about
multi-agent computations with unboundedly many agents.  For example, a
\emph{computation} in sccp is a possibly infinite reduction sequence $\gamma$
of the form $\langle \ P_0, c_0 \ \rangle \longrightarrow \langle\ P_1, c_1 \
\rangle \longrightarrow  \cdots $ with $c_0 \cleq c_1\cleq \cdots $. The
\emph{result} of $\gamma$ is $\bigjoin_{n\geq 0} c_n$, the join of all the
stores in the computation. In sccp all fair computations from a configuration
have the same result~\cite{knight:hal-00761116}. Thus, the \emph{observable
behaviour} of $P$ with initial store $c$, written $\mathcal{O}(P,c)$, is
defined as the result of any fair computation starting from $\langle P, c
\rangle.$  Now consider a setting where in addition to their sccp capabilities
in~\cite{knight:hal-00761116}, processes can also create new agents. Hence,
unboundedly many agents, say agents $1,2,\ldots$, may be created during an
infinite computation. In this case, $\mathcal{O}(P,c) \cgeq
\Dfunapp{\mathbb{N}}{\false}$, where $\mathbb{N}$ is  the set of natural
numbers, would imply that some (finite or infinite) set of agents in any fair
computation from $\langle P, c \rangle$ may reach contradictory local
information among them. Notice that from the above-mentioned  properties of
distributed spaces, the existence of a finite set of agents $H \subseteq
\mathbb{N}$ such that $\mathcal{O}(P,c) \cgeq \Dfunapp{H}{\false}$ implies
$\mathcal{O}(P,c) \cgeq \Dfunapp{\mathbb{N}}{\false}$. The converse of this
implication will be called \emph{group compactness} and we will discuss
meaningful sufficient conditions for it to hold.   

In the next sections we will describe the above spatial and epistemic notions
in more detail. 

%%% END %%%%

% Section 3
\section{Background}
\label{sec:back}
We presuppose basic knowledge of domain and order theory 
\cite{davey2002introduction,abramsky1994domain,gierz2003continuous} and use
the following notions. Let $\C$ be a poset $(\Con, \cleq)$, and let $S
\subseteq \Con$. We use $\bigjoin S$ to denote the least upper bound (or
\emph{supremum} or \emph{join}) of the elements in $S$, and  $\bigmeet S$ is
the greatest lower bound (glb) (\emph{infimum} or  \emph{meet}) of the
elements in $S$.  An element $e \in S$ is the \emph{greatest element} of $S$
iff for every element $e' \in S$, $e' \cleq e$. If such $e$ exists, we denote
it by $\textit{max}\ S$. As usual, if $S= \{ c,d \}$, $c \sqcup d$ and $c
\meet d$ represent $\bigjoin S$ and $\bigmeet S$, respectively. If $S =
\emptyset$, we denote $\bigjoin S = \true$ and $\bigmeet S = \false$. We say
that $\C$ is a \emph{complete lattice} iff each subset of $\Con$ has a
supremum in $\Con$.  The poset $\C$ is \emph{distributive} iff for every
$a,b,c \in \Con$, $a \join ( b \meet c) =  (a \join b) \meet (a \join c).$ A
non-empty set $S \subseteq \Con$ is \emph{directed} iff for every pair of
elements $x, y \in S$, there exists $z \in S$ such that $x \cleq z$ and $y
\cleq z$, or iff every \emph{finite} subset  of $S$ has an upper bound in $S$.
Also $c\in{\Con}$ is \emph{compact} iff for any  directed subset $D$ of
${\Con}$, $c \cleq \bigjoin D$ implies $c \cleq d$ for some $d\in D$. A
\emph{self-map} on $\Con$ is a function $f$ from  $\Con$ to $\Con$. Let
$(\Con, \cleq)$ be a complete lattice. The self-map $f$ on $\Con$
\emph{preserves}  the join of a set $S \subseteq \Con$  iff $f(\bigjoin S) =
\bigjoin \{ f(c)\mid c \in S \}$. A self-map that preserves the join of finite
sets is called \emph{join-homomorphism}. A self-map $f$ on $\Con$ is
\emph{monotonic}  if $a \cleq b$ implies $f(a) \cleq f(b)$.  We say that $f$
\emph{distributes} over  joins (or that $f$ \emph{preserves} joins) iff it
preserves the join of arbitrary sets. A self-map $f$ on ${\Con}$ is
\emph{continuous} iff it preserves the join of any directed set. 

\label{sec:scs}

\emph{Constraint systems}~\cite{saraswat1991semantic} are semantic structures
to specify partial information. They can be formalized as complete
lattices~\cite{de1995nondeterminism}.

\begin{definition}[Constraint Systems~\cite{de1995nondeterminism}]
\label{def:cs}
A \emph{constraint system} (cs) $\C$ is a complete lattice $(\Con, \cleq)$.
The elements of $\Con$ are called \emph{constraints}. The symbols $\sqcup$,
$\true$ and $\false$ will be used to denote the least upper bound (lub)
operation, the bottom, and the top element of $\C$.
\end{definition}

The elements of the lattice, the \emph{constraints}, represent (partial)
information. A constraint $c$ can be viewed as an \emph{assertion}. The
lattice order  $\cleq$ is meant to capture entailment of information: $c \cleq
d$, alternatively written  $d \cgeq c$, means that the assertion $d$
represents at least as much information as $c$. We think of $d \cgeq c$ as
saying that $d$ \emph{entails} $c$ or that $c$ can be \emph{derived} from $d$.
The operator $\sqcup$ represents join of information;  $c \join d$ can be
seen as an assertion stating that both $c$ and $d$ hold. We can think of
$\sqcup$ as representing conjunction of assertions.  The top element
represents the join of all, possibly inconsistent, information, hence it is
referred to as  $\false$. The bottom element  $\true$ represents \emph{empty
information}.  We say that $c$ is \emph{consistent} if $c\neq \false$,
otherwise  we say that $c$ is \emph{inconsistent.} Similarly, we say that $c$
is  consistent/inconsistent with $d$ if $c\sqcup d$ is
consistent/inconsistent. 

\emph{Constraint Frames.}
One can define a general form of implication by adapting the corresponding
notion from Heyting Algebras to cs. A \emph{Heyting implication}
$c \to d$ in our setting corresponds to the \emph{weakest constraint} 
one needs to join $c$ with to derive $d$.

\begin{definition}[Constraint Frames \cite{guzman:hal-01257113}]
\label{def:frames}
A constraint system $(\Con, \cleq)$ is said to be a \emph{constraint frame}
iff its joins distribute over arbitrary meets. More precisely,
$c \join {\bigmeet S} = \bigmeet\{ c \join e \mid e \in S\}$ for every
$c \in \Con$ and $S \subseteq \Con$.  Define
$c \imp d$ as $\bigmeet \{ e \in \Con \mid { {c} \join {e}} \cgeq {d} \}$.
\end{definition}

The following properties of Heyting implication correspond to standard logical
properties (with $\to$, $\join$, and $\cgeq$ interpreted as implication,
conjunction, and entailment).

\begin{proposition}[\cite{guzman:hal-01257113}]
\label{prop:implication}
Let  $(\Con, \cleq)$ be a constraint frame. For every $c,d,e \in \Con$ the 
following holds:
(1) \label{proof:modus-ponens} $c \join (c \to d) =  c \join d$,
(2) $(c \to d) \cleq  d$,
(3) $c \to  d = \true$ iff $c \cgeq  d$.
\end{proposition}

\section{Space and Beliefs}

The authors of~\cite{knight:hal-00761116}
extended the notion of cs to account for  distributed and multi-agent
scenarios with a finite number of agents, each having their own space for
local information and their computations. The extended structures are called
spatial cs (scs).  Here we adapt scs to reason about  possibly infinite groups
of agents.

A \emph{group} $G$ is a set of agents. Each $i \in G$ has a \emph{space}
function  $\sfun{i}: \Con \to \Con$  satisfying some structural conditions. 
Recall that constraints can be viewed as assertions. Thus given  $c \in \Con$,
we can then think of the constraint $\sfunapp{i}{c}$ as an assertion  stating
that $c$ is a piece of information residing \emph{within a space of  agent}
$i$. Some alternative \emph{epistemic} interpretations of $\sfunapp{i}{c}$ is
that it is an  assertion stating that agent $i$ \emph{believes} $c$, that $c$
holds within the space of agent $i$, or that agent $i$ \emph{interprets} $c$
as $\sfunapp{i}{c}$. All these interpretations convey the  idea that $c$ is
local or subjective to agent $i$. 

In \cite{knight:hal-00761116} scs are used to specify the spatial distribution
of information in configurations  $\langle P, c \rangle$ where $P$ is a
process and $c$ is a constraint. E.g., a reduction  $ \langle \ P,
{\sfunapp{i}{c}}\join {\sfunapp{j}{d}} \ \rangle \longrightarrow \langle\  Q,
{\sfunapp{i}{c}} \join {\sfunapp{j}{d \join e}} \ \rangle$ means that $P$ with
$c$ in the space of agent $i$ and $d$ in the space  of agent $j$ can evolve to
$Q$ while adding $e$ to the space of agent $j$. 

We now introduce the notion of space function.

\begin{definition}[Space Functions]
\label{def:scs}
A \emph{space function} over a cs $(\Con,\cleq)$ is a \emph{continuous}
self-map $f:\Con \to \Con$ s.t. for every $c,d \in \Con$
(S.1) $f(\true)=\true$,
(S.2) $f(c \join d) = f(c) \join f(d)$.
We shall use $\sfunspace{\C}$ to denote the set of all space functions over
$\C=(\Con, \cleq)$.  
\end{definition}

The assertion $f(c)$ can be viewed as saying that $c$ is in the space
represented by $f$.  Property S.1 states that having an empty local space
amounts to nothing. Property S.2  allows us to  join and distribute the
information in the space represented by $f$. 

In~\cite{knight:hal-00761116} space functions were not required to be
continuous. Nevertheless, continuity  comes naturally in the intended phenomena we wish to capture:
modelling information of possibly \emph{infinite} groups.  In fact,
in~\cite{knight:hal-00761116} scs could only have finitely many agents. 

In \cite{guzman:hal-02172415} we extended scs to allow  arbitrary, possibly infinite, sets
of agents.  A \emph{spatial cs} is a cs with a possibly infinite group of agents 
each having a space function. 

\begin{definition}[Spatial Constraint Systems]
A \emph{spatial cs (scs)} is a cs $\C = (\Con, \cleq)$ equipped with a
possibly infinite tuple $\sfuntuple = (\sfun{i})_{i \in G}$ of space functions
from $\sfunspace{\C}$.

We shall use $({\Con},\cleq,(\sfun{i})_{i \in G})$ to denote an scs with a
tuple $(\sfun{i})_{i \in G}$.  We refer to $G$ and $\sfuntuple$ as the
\emph{group of agents} and \emph{space tuple} of $\C$ and to each $\sfun{i}$
as the \emph{space function} in $\C$ of agent $i$. Subsets of $G$ are also
referred to as groups of agents (or sub-groups of $G$).
\end{definition}
Let us illustrate a simple scs that will be used throughout the paper.
\begin{example}
\label{simple:example}
The scs $({\Con},\cleq,(\sfun{i})_{i \in \{1,2\}})$  in Fig.\ref{ex:spatial}
is  given by the complete lattice $\mathbf{M}_2$ and two agents. We have $\Con
= \{ p \vee \neg p, p, \neg p, p \wedge \neg p \}$ and $c \cleq d$ iff $c$ is
a logical consequence of $d$.  The top element $\false$ is  $p \wedge \neg p$,
the bottom element $\true$ is  $p \vee \neg p$, and the constraints $p$ and $\neg p$ are
incomparable with each other. The set of agents is $\{1,2\}$ with space
functions $\sfun{1}$ and $\sfun{2}$: For agent $1$, $\sfunapp{1}{p}=\neg p$,
$\sfunapp{1}{\neg p}=p$, $\sfunapp{1}{\false} = \false$, $\sfunapp{1}{\true} =
\true$, and for agent $2$, $\sfunapp{2}{p} = \false = \sfunapp{2}{\false}$,
$\sfunapp{2}{\neg p} = \neg p$, $\sfunapp{2}{\true} = \true$.   The intuition
is that the agent $2$  sees no difference between $p$ and $\false$ while agent
$1$ interprets $\neg p$ as $p$ and vice versa. 
\end{example}
More involved examples of scs include meaningful families of structures from
logic and economics such as Kripke structures and Aumann structures (see \cite{knight:hal-00761116}). We illustrate scs with infinite groups in the next section.  
\begin{figure}[t]
\centering
\resizebox{0.33\textwidth}{!}{%
\begin{tikzpicture}[ node distance=1.3cm]
  % \tikzstyle{every node}=[font=\LARGE]
  \node[shape = circle, draw] (A) at (0,-2.5)   {$p \vee \neg p$};
  \node[shape = circle, draw] (B) at (-2.5,0)   {$p$};
  \node[shape = circle, draw] (C) at (2.5,0)    {$\neg p$};
  \node[shape = circle, draw] (D) at (0,2.5)    {$p \wedge \neg p$};

  \draw (A) to node {} (B);
  \draw (A) to node {} (C);
  \draw (B) to node {} (D);
  \draw (C) to node {} (D); 

  \draw [above,blue,thick,->,bend left]  (B) to node {$\sfun{1}$} (C);
  \draw [below,blue,thick,->,bend left]  (C) to node {$\sfun{1}$} (B);
  \draw [below,blue,thick,->,loop left]  (A) to node {$\sfun{1}$} (A);
  \draw [below,blue,thick,->,loop left]  (D) to node {$\sfun{1}$} (D);
  \draw [below,red,thick,->,loop right]  (D) to node {$\sfun{2}$} (D);
  \draw [left,red,thick,->,bend left]    (B) to node {$\sfun{2}$} (D);
  \draw [below,red,thick,->,loop right]  (A) to node {$\sfun{2}$} (A);
  \draw [below,red,thick,->,loop right]  (C) to node {$\sfun{2}$} (C);
\end{tikzpicture}
}%
\caption{Cs given by lattice $\mathbf{M}_2$ ordered by implication and space functions $\sfun{1}$ and $\sfun{2}$.}
\label{ex:spatial}
\end{figure}

\section{Extrusion and Utterances}
%Section

We can also equip each agent $i$ with an \emph{extrusion} function $\efun{i}:\Con \rightarrow \Con$. Intuitively, within a space context $\sfunapp{i}{\cdot}$, the assertion $\efunapp{i}{c}$ specifies that $c$ must be posted outside  of agent $i$'s space.  This is captured by requiring the \emph{extrusion} axiom (E.1) \( \ \sfunapp{i}{\efunapp{i}{c}} = c. \) In other words, we view \emph{extrusion/utterance} as the right inverse of \emph{space/belief} (and thus space/belief as the left inverse of extrusion/utterance). 

\begin{definition}[Extrusion]\label{scse} Given an scs $({\Con},\cleq,(\sfun{i})_{i \in G})$, we say that $\efun{i}$ is an extrusion function for the space $\sfun{i}$ iff  $\efun{i}$ is a right inverse of $\sfun{i}$, i.e., 
 iff \( \ \sfunapp{i}{\efunapp{i}{c}} = c. \) 
\qed\end{definition}

From the above definitions it follows that $\sfunapp{i}{c \sqcup \efunapp{i}{d}} =  \sfunapp{i}{c } \sqcup d.$ From a spatial point of view, agent $i$ \emph{extrudes} $d$ from its local space. From an epistemic view this can be seen as an agent $i$ that believes $c$ and \emph{utters} $d$ to the outside world. If $d$ is inconsistent with 
$c$, i.e., $c\sqcup d = \false$, we can see the utterance as an intentional \emph{lie} by agent $i$: The agent $i$ utters an assertion inconsistent with their own beliefs.  

\begin{example}\label{extrusion-example} Let $e = \sfunapp{i}{c \sqcup  \efunapp{i}{\sfunapp{j}{a}} } \sqcup  \sfunapp{j}{d}$. The constraint $e$
specifies that agent $i$ has $c$ and wishes to transmit, via extrusion, $a$ addressed to agent $j$. Agent $j$ has $d$ in their space. Indeed, with the help of E.1 and S.2, we can derive $e \sqsupseteq  \sfunapp{j}{d \sqcup a}$ thus stating that $e$ entails that $a$ will be in the space of $j$.  
\end{example}

\paragraph{The Extrusion Problem.} A legitimate question is: Given space $\sfun{i}$ can we derive an extrusion function $\efun{i}$ for it ? From set theory we know that there is an extrusion function (i.e., a right inverse) $\efun{i}$ for $\sfun{i}$  iff  $\sfun{i}$ is \emph{surjective}.  Recall that the \emph{pre-image} of $y \in Y$ under $f : X \rightarrow Y$ is the set $f^{-1}(y) = \{ x \in X \mid y = f(x) \}$. Thus the extrusion $\efun{i}$ can be defined as a function, called \emph{choice} function, that maps each element $c$ to some element from the pre-image of $c$ under $\sfun{i}.$

The existence of the above-mentioned choice function assumes the \emph{Axiom of Choice}.   The next  proposition from \cite{guzman:hal-01257113} gives some constructive extrusion functions. It also identifies a distinctive property of space functions for which a right inverse exists. 

\index{$\sfunc{\cdot}^{-1}$, pre-image under $\sfunc{\cdot}$ }
\begin{proposition}[\cite{guzman:hal-01257113}]\label{extrusion-prop} Let $f$ be  a space function over $({\Con},\cleq)$. Then
\begin{enumerate}
\item If  $f(\false)\neq \false$ then $f$ does not have any right inverse. 
\item If  $f$ is surjective  then $g: c \mapsto \bigsqcup f(c)^{-1}$ is a right inverse of $f$ that preserves arbitrary infima. 
\item If  $f$ is surjective and preserves arbitrary infima then $h: c \mapsto \bigsqcap f(c)^{-1}$ is a right inverse of $f$ that preserves arbitrary suprema.
%\item If $\sfunc{\cdot}_i$ preserves arbitrary infima then $\efunc{_i}: c \mapsto \bigsqcap \sfunc{c}_i^{-1}$ is a right inverse of $\sfunc{\cdot}_i$ and a normal self-map. 
\end{enumerate}
\end{proposition}

The following example illustrates an application of Prop.\ref{extrusion-prop} to obtain an extrusion function for the space function $\sfun{1}$ from
Ex.\ref{simple:example}. Notice that the space function $\sfun{2}$  from Ex.\ref{simple:example} is not surjective thus it 
does not have an extrusion function.

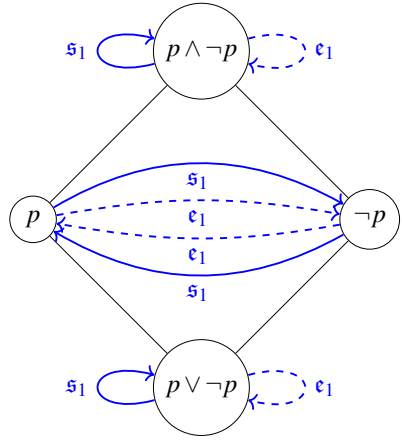
\begin{figure}[t]
\centering
\resizebox{0.33\textwidth}{!}{%
\begin{tikzpicture}[ node distance=1.3cm]
 \tikzstyle{every node}=[font=\normalsize]
  \node[shape = circle, draw] (A) at (0,-2.5)   {$p \vee \neg p$};
  \node[shape = circle, draw] (B) at (-2.5,0)   {$p$};
  \node[shape = circle, draw] (C) at (2.5,0)    {$\neg p$};
  \node[shape = circle, draw] (D) at (0,2.5)    {$p \wedge \neg p$};

  \draw (A) to node {} (B);
  \draw (A) to node {} (C);
  \draw (B) to node {} (D);
  \draw (C) to node {} (D); 

  \draw [below,blue,thick,->,bend left]  (B) to node {$\sfun{1}$} (C);
  \draw [below,blue,thick,->,bend left]  (C) to node {$\sfun{1}$} (B);
  \draw [below,blue,thick,->,loop left]  (A) to node {$\sfun{1}$} (A);
  \draw [below,blue,thick,->,loop left]  (D) to node {$\sfun{1}$} (D);
  
  \draw [below,blue,dashed,->,loop right,thick]  (D) to node {$\efun{1}$} (D);
  \draw [below,blue,dashed,->,loop right,thick]  (A) to node {$\efun{1}$} (A);
  
     \draw [below,blue,dashed,->,thick,out=190,in=-10]  (C) to node {$\efun{1}$} (B);
  \draw [below,blue,dashed,->,thick,out=10, in=170]  (B) to node {$\efun{1}$} (C);
  
\end{tikzpicture}
}%
\caption{Cs given by lattice $\mathbf{M}_2$ ordered by implication and the space function $\sfun{1}$ with extrusion $\efun{1}$.}
\label{ex:extrusion-graph}
\end{figure}

\begin{example}\label{extrusion-graph-example} Fig.\ref{ex:extrusion-graph} shows an extrusion function for the space function $\sfun{1}$ in
Ex.\ref{simple:example}. This extrusion function  can be obtained by applying Prop.\ref{extrusion-prop}.2.
\end{example}

% Section 4
\section{Groups and Distributed Knowledge}
\label{sec:dist-info}

In \cite{guzman:hal-02172415}  we introduced the notion of collective information of a
group of agents. Roughly speaking,  the
\emph{distributed (or collective) information} of a group $I$ is the join of
each piece of information that resides in the space of \emph{some} $i \in I$. 
The distributed information of $I$ w.r.t. $c$ is the distributive information
of $I$ that can be derived from $c$.  We wish to formalize whether
a given $e$ can be derived from the collective information of the group $I$
w.r.t. $c$. 

The following examples, which we will use throughout this section, illustrate the above intuition. 
\begin{example}
\label{ds-example}
Consider an scs $({\Con},\cleq, (\sfun{i})_{i \in G})$ where $G = \nmat$
and $(\Con, \cleq)$ is a constraint frame. Let $c \defsymbol \sfunapp{1}{a}
\join \sfunapp{2}{a \to b} \join \sfunapp{3}{b \to e}$. The constraint
$c$ specifies the situation where $a, a \to b$ and $b \to e$ are in the spaces
of agent $1$, $2$ and $3$, respectively. Neither agent necessarily holds $e$ in their
space in $c$. Nevertheless, the information $e$ can be derived from the
collective information of the three agents w.r.t. $c$, since from
Prop.\ref{prop:implication} we have $a \join (a \to b) \join (b \to e) \cgeq
e.$   Let us now consider an example with infinitely many agents.  	Let $c'
\defsymbol \bigjoin_{i \in \nmat} \sfunapp{i}{a_i}$ for some increasing
chain $a_0 \cleq a_1 \cleq \ldots.$  Take $e'$ s.t. $e' \cleq \bigjoin_{i \in
\nmat} {a_i}$. Notice that unless $e'$ is compact (see Section
\ref{sec:back}), it may be the case that no agent $i \in  \nmat$ holds
$e'$ in their space; e.g., if  $e' \cg a_i $ for any $i \in \nmat$. Yet,
from our assumption, $e'$ can be derived from the collective information
w.r.t. $c'$ of all the agents in $\nmat,$ i.e., $\bigjoin_{i \in
\nmat} {a_i}$.
\end{example}

The above example may suggest that the distributed information can be obtained
by joining individual local information derived from $c$. Individual
information of an agent $i$ can be characterized as the $i$-projection of $c$
defined thus: 

\begin{definition}[Agent and Join Projections~\cite{guzman:hal-02172415}]
\label{scsp}
Let  $\C = ({\Con},\cleq,(\sfun{i})_{i \in G})$ be an scs. Given $i\in G$, the
$i$-agent projection of  $c \in \Con$ is defined as $\pfunapp{i}{c} \defsymbol
\bigjoin \{ e \ | \ c  \cgeq \sfunapp{i}{e} \}.$   We say that $e$ is
$i$-agent derivable from $c$ iff $\pfunapp{i}{c} \cgeq e$.
Given $I \subseteq G$ the $I$-join projection of a group $I$ of $c$ is defined
as $\pfunapp{I}{c} \defsymbol \bigjoin\{ \pfunapp{i}{c} \ | \ i\in I \}$.
We say that $e$ is $I$-join derivable from $c$ iff $\pfunapp{I}{c} \cgeq e$. 
\end{definition}

The $i$-projection of an agent $i$ of  $c$ naturally represents the join of 
all the information of agent $i$ in $c$.   It turns out that projections are extrusion functions: If $\sfun{i}$  admits extrusion then $\pfun{i}$ is an extrusion function for the space $\sfun{i}$ (see Def.\ref{scse}).
More precisely,

\begin{proposition}[Projection as extrusion]  If $\sfun{i}$  is surjective then  $ \sfunapp{i}{\pfunapp{i}{c} } = c$ for every $c\in  \Con.$ 
\end{proposition} 

The $I$-join projection of group $I$ joins individual $i$-projections of $c$ for $i\in I$. This projection can be
used as a sound mechanism for reasoning about distributed-information: If $e$
is $I$-join derivable from $c$ then it follows from the
distributed-information of $I$ w.r.t. $c$. 
\begin{example}
\label{ds-example-join}
Let $c$ be as in Ex.\ref{ds-example}.  We have $\pfunapp{1}{c} \cgeq a$,
$\pfunapp{2}{c} \cgeq (a \to b)$, $\pfunapp{3}{c}  \cgeq  (b \to e)$. Indeed
$e$ is $I$-join derivable from $c$ since $\pfunapp{\{1,2,3\}}{c} =
\pfunapp{1}{c} \join \pfunapp{2}{c}\join  \pfunapp{3}{c} \cgeq e$. Similarly
we conclude that $e'$ is $I$-join derivable
from $c'$  in Ex.\ref{ds-example} since $\pfunapp{\nmat}{c'} = \bigjoin_{i \in
\nmat}\pfunapp{i}{c}\cgeq \bigjoin_{i \in \nmat} {a_i} \cgeq e'$.
\end{example}

Nevertheless, $I$-join projections do not provide a complete mechanism for
reasoning about distributed information as illustrated below. 

\begin{example}
\label{ds:counter-example}
Let $d \defsymbol \sfunapp{1}{b} \meet \sfunapp{2}{b}$. Recall that we think
of $\join$ and $\meet$ as conjunction and disjunction of assertions: $d$
specifies that $b$ is present in the space of agent $1$ or in the space of
agent $2$ though not exactly in which one. Thus from $d$ we should be able to
conclude that $b$ belongs to the space of \emph{some} agent in $\{ 1, 2 \}$.
Nevertheless, in general $b$ is not $I$-join derivable from $d$ since  from
$\pfunapp{\{1,2\}}{d} = \pfunapp{1}{d} \join \pfunapp{2}{d}$ we cannot, in
general, derive $b$.  To see this consider the scs  in
Fig.\ref{projection:counter-example} and take $b=\neg p$. We have
$\pfunapp{\{1,2\}}{d} = \pfunapp{1}{d} \join \pfunapp{2}{d}=\true \join
\true=\true \not\cgeq b$. One can generalize the example to infinitely many
agents: Consider the scs in Ex.\ref{ds-example}. Let $d' \defsymbol
\bigmeet_{i \in \nmat}\sfunapp{i}{b'}$. We should be able to conclude from
$d'$ that $b'$ is in  the space of \emph{some} agent in $\nmat$ but, in
general, $b'$ is not $\nmat$-join derivable from $d'$.
\end{example}

\subsection{Distributed Spaces}
\label{ssec:properties-ds}

In the previous section we illustrated that the $I$-join projection of $c$,
$\pfunapp{I}{c}$,  the join of individual projections, may not project all
distributed information of a group $I$.  To solve this problem we shall
develop the notion of $I$-group projection of  $c$, written as
$\Pfunapp{I}{c}$. To do this we shall first define a space function $\Dfun{I}$
called the  distributed space of group $I$. The function $\Dfun{I}$  can be
thought of as a virtual space including all the information that  can be in 
the space of  a member of $I.$  We shall then define an $I$-projection
$\Pfun{I}$ in terms of $\Dfun{I}$ much like $\pfun{i}$ is defined in terms of
$\sfun{i}$.

Recall that $\mathcal{S}(\C)$ denotes the set of all space functions over a cs
$\C$. For notational convenience, we shall use $(f_I)_{I \subseteq G}$ to
denote the tuple $(f_I)_{I \in \pcal(G)}$ of elements of $\mathcal{S}(\C)$. 

\emph{Set of Space Functions}. We begin by introducing a new partial order
induced by $\C$. The set of space functions ordered point-wise.  

\begin{definition}[Space Functions Order]
Let $\C= ({\Con},\cleq,(\sfun{i})_{i \in G})$ be an scs. Given
$f,g \in \mathcal{S}(\C)$, define $f  \fleq  g$ iff $f(c) \cleq g(c)$
for every $c\in\Con$. We shall use $\Cs$ to denote the partial order
$(\mathcal{S}(\C),\fleq )$; the set of all space functions ordered by
$\fleq$.
\end{definition}

A very important fact for the design of our structure is that the set of space
functions $\mathcal{S}(\C)$ can be made into a complete lattice.

\begin{lemma}[\cite{guzman:hal-02172415}]
\label{lemma:closure-space}
Let $\C= ({\Con},\cleq,(\sfun{i})_{i \in G})$ be an scs.
Then $\Cs$ is a complete lattice. 
\end{lemma}

\subsection{Distributed Spaces as Maximum Spaces}

Let us consider the lattice of space functions  $\Cs = (\mathcal{S}(\C),\fleq
)$. Suppose that $f$ and $g$ are space functions in $\Cs$  with $f  \fleq  g$.
Intuitively, every piece of information $c$ in the space represented by $g$ is
also in the space represented by $f$ since  $f(c) \cleq g(c)$ for every
$c\in\Con$.  This can be interpreted as saying that the space represented by
$g$ is included in the space represented by $f$; in other words the bigger the
space, the smaller the function that represents it in the lattice $\Cs$. 

Following the above intuition, the order relation $\fleq $ of $\Cs$ represents
(reverse) space inclusion and  the join and meet operations in $\Cs$
represent intersection and union of spaces. The biggest and the smallest
spaces are represented by the bottom  and the top  elements  of the lattice
$\Cs$, here called $\lambda_\bot$ and $\lambda_\top$ and defined as follows.

\begin{definition}[Top and Bottom Spaces]
\label{lambda:def}
For every $c \in \Con$, define  $\lambda_\bot(c) \defsymbol \true$, 
$\lambda_\top(c)\defsymbol \true  \mbox{ if } c = \true$ and
$\lambda_\top(c) \defsymbol \false  \mbox{ if } c \neq \true$.
\end{definition}

The distributed space  $\Dfun{I}$ of a group $I$ can be viewed as the function
that represents the smallest space that includes all the local information of
the agents in $I$.  From the above intuition, $\Dfun{I}$ should be the
\emph{greatest space function} below the space functions of the agents in $I$.
The existence of such a function follows from completeness of
$(\mathcal{S}(\C),\fleq )$ (Lemma \ref{lemma:closure-space}).  

\begin{definition}[Distributed Spaces~\cite{guzman:hal-02172415}]
\label{ds:def}
Let  $\C$ be an scs $({\Con},\cleq,(\sfun{i})_{i \in G}).$ The
\emph{distributed spaces} of  $\C$ is given by $\Dfuntuple = (\Dfun{I})_{I
\subseteq G}$ where
\(
\Dfun{I} \defsymbol {\it max} \{ f \in \mathcal{S}(\C) \ |\ f  \fleq \sfun{i} 
\mbox{ for every } i \in I  \}.
\)
We shall say that $e$ is distributed  among $I\subseteq G$ w.r.t. $c$ 
iff $c \cgeq \Dfunapp{I}{e}$. We shall refer to each $\Dfun{I}$ as the
\emph{(distributed) space} of the group $I$. 
\end{definition}

\begin{figure}[t]
\centering
\begin{subfigure}[t]{0.4\textwidth}
\centering
\resizebox{4.5cm}{!}{%
\begin{tikzpicture}[ node distance=1.3cm]
  \tikzstyle{every node}=[font=\normalsize]
  \node[shape = circle, draw] (A) at (0,-3.5)   {$p \vee \neg p$};
  \node[shape = circle, draw] (B) at (-3.5,0)   {$p$};
  \node[shape = circle, draw] (C) at (3.5,0)    {$\neg p$};
  \node[shape = circle, draw] (D) at (0,3.5)    {$p \wedge \neg p$};

  \draw (A) to node {} (B);
  \draw (A) to node {} (C);
  \draw (B) to node {} (D);
  \draw (C) to node {} (D);
  \draw [below,blue,thick,->,bend left]  (B) to node {$\sfun{1}$} (C);
  \draw [below,blue,thick,->,bend left]  (C) to node {$\sfun{1}$} (B);
  \draw [below,blue,thick,->,loop left]  (A) to node {$\sfun{1}$} (A);
  \draw [below,blue,thick,->,loop left]  (D) to node {$\sfun{1}$} (D);
  \draw [below,red,thick,->,loop right]  (D) to node {$\sfun{2}$} (D);
  \draw [left,red,thick,->,bend left]    (B) to node {$\sfun{2}$} (D);
  \draw [below,red,thick,->,loop right]  (A) to node {$\sfun{2}$} (A);
  \draw [below,red,thick,->,loop right]  (C) to node {$\sfun{2}$} (C);

  \draw [below,blue,dashed,loop below,->,thick]      (A) to node {$\pfun{1}$} (A);
  \draw [below,blue,dashed,loop below,->,thick]      (D) to node {$\pfun{1}$} (D);
  \draw [below,blue,dashed,->,thick,out=190,in=-10]  (C) to node {$\pfun{1}$} (B);
  \draw [below,blue,dashed,->,thick,out=10, in=170]  (B) to node {$\pfun{1}$} (C);

  \draw [below,loop above,red,dashed,->,thick]  (A) to node {$\pfun{2}$} (A);
  \draw [below,loop above,red,dashed,->,thick]  (D) to node {$\pfun{2}$} (D);
  \draw [below,loop above,red,dashed,->,thick]  (C) to node {$\pfun{2}$} (C);
  \draw [left,bend right,red,dashed,->,thick]   (B) to node {$\pfun{2}$} (A);
\end{tikzpicture}
}%
\caption{Projections $\pfun{1}$ and $\pfun{2}$ given $\sfun{1}$  and $\sfun{2}$.}
\label{projection:counter-example}
\end{subfigure}
~
\begin{subfigure}[t]{0.4\textwidth}
\centering
\resizebox{4.5cm}{!}{%
\begin{tikzpicture}[ node distance=1.3cm]
  \tikzstyle{every node}=[font=\Large]
  \node[shape = circle, draw] (A) at (0,-3)   {$p \vee \neg p$};
  \node[shape = circle, draw] (B) at (-3,0)   {$p$};
  \node[shape = circle, draw] (C) at (3,0)    {$\neg p$};
  \node[shape = circle, draw] (D) at (0,3)    {$p \wedge \neg p$};

  \draw (A) to node {} (B);
  \draw (A) to node {} (C);
  \draw (B) to node {} (D);
  \draw (C) to node {} (D);
  \draw [above,blue,thick,->,bend left]  (B) to node {$\sfun{1}$} (C);
  \draw [below,blue,thick,->,bend left]  (C) to node {$\sfun{1}$} (B);
  \draw [below,blue,thick,->,loop left]  (A) to node {$\sfun{1}$} (A);
  \draw [below,blue,thick,->,loop left]  (D) to node {$\sfun{1}$} (D);
  \draw [below,red,thick,->,loop right]  (D) to node {$\sfun{2}$} (D);
  \draw [left,red,thick,->,bend left]    (B) to node {$\sfun{2}$} (D);
  \draw [below,red,thick,->,loop right]  (A) to node {$\sfun{2}$} (A);
  \draw [below,red,thick,->,loop right]  (C) to node {$\sfun{2}$} (C);

  \draw [below,purple,dashed,loop below,->,thick]      (A) to node {$\Dfun{I}$} (A);
  \draw [right,purple,dashed,bend left,->,thick]       (D) to node {$\Dfun{I}$} (C);
  \draw [right,purple,dashed,bend left,->,thick]       (C) to node {$\Dfun{I}$} (A);
  \draw [above,purple,dashed,->,thick]                 (B) to node {$\Dfun{I}$} (C);
\end{tikzpicture}
}%
\caption{$\Dfun{I}$ with $I=\{1,2\}$ given 
$\sfun{1}$ and $\sfun{2}$.}
\label{fig:difficult-small}
\end{subfigure}
\caption{Projections (a) and Distributed Space function (b) over lattice $\mathbf{M}_2$.}
\label{fig:projections-delta}
\end{figure}
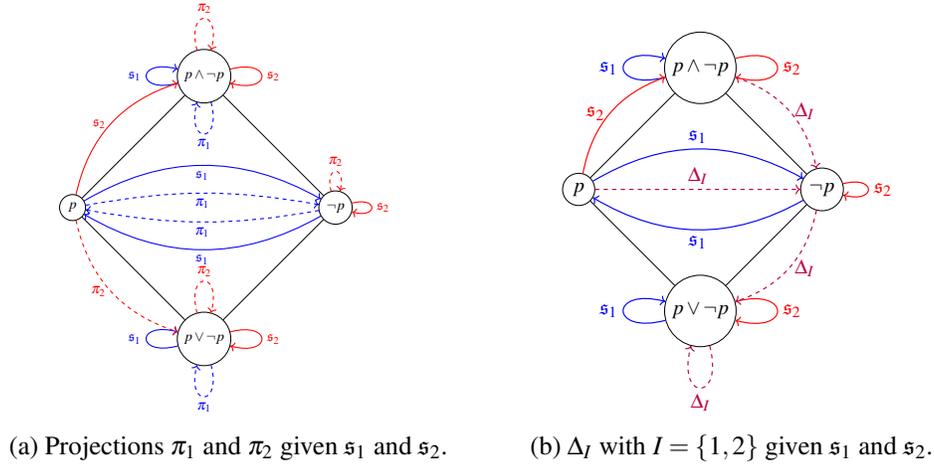

It follows from Lemma \ref{lemma:closure-space} that $\Dfun{I}=\bigmeet\{ 
\sfun{i} \ | \ i \in I  \}$ (where $\bigmeet$ is the  meet  in the complete
lattice  $(\mathcal{S}(\C),\fleq )$).  Fig.\ref{fig:difficult-small} 
illustrates an scs and its distributed space $\Dfun{\{1,2\}}.$

\emph{Compositionality}. Distributed spaces have pleasant compositional
properties. They capture the intuition that the \emph{distributed information}
of a  group $I$ can be obtained from the the distributive information of its
subgroups.  

\begin{theorem}[\cite{guzman:hal-02172415}]
\label{thm:comp}
Let $(\Dfun{I})_{I \subseteq G}$ be the distributed spaces of an scs
$(\Con,\cleq,(\sfun{i})_{i\in G})$. Suppose that $K,J \subseteq I\subseteq G$.
(1) $\Dfun{I} = \lambda_\top $ if  $I = \emptyset $, 
(2) $\Dfun{I}=\sfun{i}\ $  if $I=\{ i \}$,   
(3) $\Dfunapp{J}{a} \join \Dfunapp{K}{b} \cgeq  \Dfunapp{I}{a \join b}$, and
(4) $\Dfunapp{J}{a} \join \Dfunapp{K}{a \to c} \cgeq  \Dfunapp{I}{c}$ if
$({\Con},\cleq)$ is a constraint frame. 
\end{theorem}

Recall that $\lambda_\top$  corresponds to the empty space (see
Def.\ref{lambda:def}). The first property realizes the intuition that the
empty subgroup $\emptyset $ \emph{does not} have any information whatsoever
distributed w.r.t. a consistent $c$: for if $c \cgeq \Dfunapp{\emptyset}{e}$
and $c \neq \false$  then $e = \true$. Intuitively, the second property says
that the function $\Dfun{I}$ for the group of one agent must be the agent's
space function. The third property states that a group can join the
information of its subgroups. The last property uses constraint implication,
hence the constraint frame condition, to express that by joining the
information $a$ and $a \to c$ of their subgroups, the group $I$ can obtain
$c$.  

Let us illustrate how to derive information of a group from smaller ones using
Thm.\ref{thm:comp}. 

\begin{example}
Let $c = \sfunapp{1}{a} \join \sfunapp{2}{a \to b} \join \sfunapp{3}{b \to e}$
as in Ex.\ref{ds-example}.  We want to prove that $e$ is distributed among
$I=\{1,2,3\}$ w.r.t. $c$, i.e., $c \cgeq \Dfunapp{\{1,2,3\}}{e}$.  Using
Properties 2 and 4 in Thm.\ref{thm:comp} we obtain $ c \cgeq \sfunapp{1}{a}
\join \sfunapp{2}{a \to b} =   \Dfunapp{\{1\}}{a} \join \Dfunapp{\{2\}}{a \to
b} \cgeq  \Dfunapp{\{1,2\}}{b}$, and then	$c\cgeq  \Dfunapp{\{1,2\}}{b}
\join \sfunapp{3}{b \to e}  =   \Dfunapp{\{1,2\}}{b} \join 
\Dfunapp{\{3\}}{b \to e} \cgeq  \Dfunapp{\{1,2,3\}}{e} $ as wanted. 
\end{example}

\begin{remark}[Continuity]
\label{remark:cont}
The example with infinitely many agents in Ex.\ref{ds-example} illustrates
well why we require our spaces to be continuous in the presence of possibly
infinite groups. Clearly $c' = \bigjoin_{i \in \nmat} \sfunapp{i}{a_i} \cgeq
\bigjoin_{i \in \nmat}\Dfunapp{\nmat}{a_i}$. By continuity, $ \bigjoin_{i \in
\nmat}\Dfunapp{\nmat}{a_i} = \Dfunapp{\nmat}{\bigjoin_{i \in {\nmat}}a_i}$
which indeed captures the idea that each $a_i$ is in the distributed space
$\Dfun{\nmat}$.
\end{remark}

We conclude this subsection with an important family of scs from 
mathematical economics: Aumann structures. We illustrate that the notion of 
distributed knowledge in these structures is an instance of a distributed 
space.

\begin{example}
\label{aumann:example}
\emph{Aumann Constraint Systems.} Aumann structures~\cite{halpern2004reasoning}
are an  \emph{event-based} approach to modelling knowledge.  An Aumann
structure is a tuple $\acal = (S, \pcal_1, \dots, \pcal_n)$ where $S$ is a set
of states and each $\pcal_i$ is a partition on $S$ for agent $i$.  The
partitions are called \emph{information sets}.  If two states $t$ and $u$ are
in the same information set for agent $i$, it means that in state $t$ agent
$i$ considers state $u$ possible, and vice versa.  An \emph{event} in an
Aumann structure is any subset of $S$.  Event $e$ holds at state $t$ if $t\in
e$.   The set $\pcal_i(s)$ denotes the information set of $\pcal_i$ containing
$s$.  The event of \emph{agent $i$ knowing $e$} is defined as  $\Kop{i}{e} =
\{ s\in S\ |\ \pcal_i(s) \subseteq e\}$, and  the \emph{distributed knowledge
of an event $e$ among the agents in a group $I$} is defined as $\Gop{I}{e} =
\{s \in S\ |\ \bigcap_{i\in I} \pcal_i(s) \subseteq e\}$.

An Aumann structure can be seen as a spatial constraint system
$\mathbf{C}(\acal)$ with events as constraints, i.e., $\Con = \{e\ |\ e \mbox{
is an event in } \acal\}$, and for every $e_1, e_2 \in \Con$, $e_1 \cleq e_2$
iff $e_2 \subseteq e_1$. The operators join $(\join)$ and meet $(\meet)$ are
intersection $(\cap)$ and union $(\cup)$ of events, respectively; $\true = S$
and $\false = \emptyset$. The space functions are the knowledge operators,
i.e., $\sfunapp{i}{c}= \Kop{i}{c}$.  From these definitions and since meets
are unions one can easily verify  that $\Dfunapp{I}{c}=\Gop{I}{c}$ which shows
the correspondence between distributed information and distributed knowledge.
\end{example} 

\subsection{Group Projections}

As promised in Section \ref{ssec:properties-ds} we now give a definition of
\emph{Group Projection}. The function $\Pfunapp{I}{c}$ extracts exactly all
information that the group $I$ may have distributed w.r.t. $c$.

\begin{definition}[Group Projection~\cite{guzman:hal-02172415}]
\label{group-projection}
Let  $(\Dfun{I})_{I \subseteq G}$ be the distributed spaces of an scs $\C =
({\Con},\cleq,(\sfun{i})_{i \in G}).$  Given the set $I \subseteq G$, the
$I$-group projection of  $c \in \Con$ is defined as $\Pfunapp{I}{c} \defsymbol
\bigsqcup \{ e \ | \ c  \cgeq \Dfunapp{I}{e} \}$. We say that $e$ is $I$-group
derivable from $c$ iff $\Pfunapp{I}{c}\cgeq e$. 
\end{definition}

Much like space functions and agent projections,  group projections and
distributed spaces  also form a pleasant correspondence: a Galois connection
\cite{davey2002introduction}. 

\begin{proposition}[\cite{guzman:hal-02172415}]
\label{projection:prop}
Let  $(\Dfun{I})_{I \subseteq G}$ be the distributed spaces of 
$\C = ({\Con},\cleq,(\sfun{i})_{i \in G})$. For every $c,e\in \Con$,
(1) $ c \cgeq \Dfunapp{I}{e} \  \mbox{   iff   } \  \Pfunapp{I}{c} \cgeq e $,
(2) $\Pfunapp{I}{c}\cgeq  \Pfunapp{J}{c}$ if $J \subseteq I$, and (3) $\Pfunapp{I}{c}\cgeq  \pfunapp{I}{c}.$
\end{proposition}

The first property in Prop.\ref{projection:prop}, a Galois connection, states
that we can conclude from $c$ that $e$ is in the distributed space of $I$
exactly when $e$ is $I$-group derivable from $c$. The second says that the
bigger the  group, the bigger the projection. The last property says that
whatever is $I$-join derivable is $I$-group derivable, although the opposite 
is not true as shown in Ex.\ref{ds:counter-example}.

\subsection{Group Compactness}

Suppose that an \emph{infinite} group of agents $I$ can derive $e$ from $c$
(i.e.,  $c \cgeq \Dfunapp{I}{e}$).  A legitimate question is whether there
exists a \emph{finite} sub-group $J$ of agents from $I$ that can also derive
$e$ from $c$. The following theorem provides a positive answer to this
question provided that $e$ is a compact element
and $I$-join derivable from $c$.

\begin{theorem}[Group Compactness~\cite{guzman:hal-02172415}]
\label{thm:compact}
Let  $(\Dfun{I})_{I \subseteq G}$ be the distributed spaces of an scs  $\C =
({\Con},\cleq,(\sfun{i})_{i \in G}).$ Suppose that  $c \cgeq \Dfunapp{I}{e}.$
If $e$ is compact and $I$-join derivable from $c$ then there exists a finite
set $J \subseteq I$ such that $c \cgeq \Dfunapp{J}{e}$.
\end{theorem}

We conclude this section with the following example of group compactness. 

\begin{example}
Consider the example with infinitely many agents in Ex.\ref{ds-example}. We
have $c' = \bigjoin_{i \in \nmat} \sfunapp{i}{a_i}$ for some increasing chain
$a_0 \cleq a_1 \cleq \ldots$  and  $e'$ s.t. $e' \cleq \bigjoin_{i \in \nmat}
{a_i}$. Notice that $c' \cgeq \Dfunapp{\nmat}{e'}$ and 
$\pfunapp{\nmat}{c'}\cgeq e'$. Hence $e'$ is $\nmat$-join derivable from $c'$.
If $e'$ is compact, by Thm.\ref{thm:compact} there must be a finite subset
$J\subseteq \nmat$ such that  $c' \cgeq \Dfunapp{J}{e'}$.
\end{example}

% Section 5
\section{Computing Distributed Information}
\label{algorithms:ds}

Let us consider a \emph{finite} scs $\C = ({\Con},\cleq,(\sfun{i})_{i \in G})$
with distributed spaces $(\Dfun{I})_{I \subseteq G}$.   By finite scs we mean
that $\Con$ and $G$ are finite sets.  Let us consider the problem of computing
$\Dfun{I}$: Given a set  $\{ \sfun{i} \}_{i \in I}$ of space functions, we
wish to find the greatest space function $f$ such that $f \cleq  \sfun{i}
\mbox{ for all } i \in I$ (see Def.\ref{ds:def}).

Because of the finiteness assumption, the above problem can be rephrased in
simpler terms: \emph{Given a finite lattice $L$ and a finite set $S$ of
join-homomorphisms on $L$, find the greatest join-homomorphism below all the
elements of $S$}. Even in small lattices with four elements and two space
functions, finding such greatest function may not be immediate, e.g., for
$S=\{ \sfun{1},\sfun{2} \}$ and the lattice in Fig.\ref{ex:spatial} the answer
is given Fig.\ref{fig:difficult-small}.

A \emph{brute force} approach would be to compute $\Dfunapp{I}{c}$ by generating the set $\{ f (c) \ | \ f \in \sfunspace{\C} \mbox{
and } f \cleq  \sfun{i} \mbox{ for all } i \in I\}$ and taking its join. This
approach works since $(\bigjoin S)(c) = \bigjoin\{ f(c) | f \in S\}.$ However, the
number of such functions in $\sfunspace{\C}$  can be at least factorial in the
size of $\Con$. For distributive lattices, the size of $\sfunspace{\C}$ can be
non-polynomial in the size of $\Con$.

\begin{proposition}[\cite{guzman:hal-02172415}]
\label{pr:size-sf}
For every $n\geq 2$, there exists a lattice $\C = ({\Con},\cleq)$ such that
$|\sfunspace{\C}| \geq (n - 2)!$ and $n = |\Con\, |$. For every $n \geq 1$, 
there exists a distributed lattice $\C = ({\Con},\cleq)$ such that
$|\sfunspace{\C}| \geq n^{\log_2 n}$  and  $n = |\Con\, |$.
\end{proposition} 

Nevertheless, we can exploit order theoretical results and
compositional properties of distributive spaces to compute $\Dfun{I}$ in
 polynomial time in the size of $\Con$. 
 
%, and a top-down approximation
%method using the function $\dapproxfunapp{I}{c}
%\defsymbol\bigmeet\{  \sfun{i}(c) \ | \ i \in I  \}$ for each $c \in \Con$ as
%a suitable upper bound for arbitrary lattices. 

%
%\begin{lemma}
%\label{lemma:delta-ast}
%Let  $(\Dfun{I})_{I \subseteq G}$ be the distributed spaces of a finite scs
%$\C = ({\Con},\cleq,(\sfun{i})_{i \in G})$.
%Suppose that $({\Con},\cleq)$ is a constraint frame.
%Let  $\dplusfun{I}:\Con \to \Con$, with $I\subseteq G$, be the function   
%$\dplusfunapp{I}{c} \defsymbol \bigmeet\{ \bigjoin_{i \in I}\sfunapp{i}{a_i}
%\ | \ (a_i)_{i \in I} \in \tuplespace{\Con}{I} \mbox { and }\
%\bigjoin_{i \in I}{a_i} \cgeq c \}$. Then $\Dfun{I} = \dplusfun{I}$.
%\end{lemma}

\begin{theorem}[\cite{guzman:hal-02172415}]
\label{thm:comp-algo} 
Suppose that $({\Con},\cleq)$ is a distributed lattice. Let $J$ and $K$ be two sets such that $I = J \cup
K.$ Then the following equalities hold:   
\begin{eqnarray}
  1.  \ \Dfunapp{I}{c} & = & \bigmeet\{ \Dfunapp{J}{a} \join 
  \Dfunapp{K}{b } \ | \ a,b \in \Con \ \mbox{ and } \ a \join b \cgeq c \}. \\
  2. \ \Dfunapp{I}{c}& = &  \bigmeet\{ \Dfunapp{J}{a} \join 
  \Dfunapp{K}{a \to c } \ | \ a \in \Con \}. \\
  3.  \ \Dfunapp{I}{c}& = &  \bigmeet\{ \Dfunapp{J}{a} \join
  \Dfunapp{K}{a \to c }  \ | \ a \in \Con \  \mbox{ and } \ a \cleq c \}.
\end{eqnarray}
\end{theorem}
The above theorem  characterizes the information of a group from that of its subgroups. 
It bears witness to the inherent compositional nature of our
notion of distributed space, and realizes the intuition that by joining the
information $a$ and $a \to c$ of their subgroups, the group $I$ can obtain $c$.
This compositional nature is exploited by the algorithms below.

Given a finite scs $\C = ({\Con},\cleq,(\sfun{i})_{i \in G})$, the recursive function \textsc{DeltaPart3}$(I,c)$
in Algorithm \ref{alg:Delta} computes $\Dfunapp{I}{c}$ for any given $c$ in $Con$. Its correctness,   assuming that
$({\Con},\cleq)$ is a distributed lattice, follows
from Thm.\ref{thm:comp-algo}(3). Termination follows from the finiteness of
$\C$ and the fact the sets $J$ and $K$ in the recursive calls form a partition
of $I$. Notice that we select a partition (in halves) rather than any two sets
$K,J$ satisfying  the condition $I = J \cup K$ to avoid significant
recalculation. 

\begin{algorithm}
\caption{Function \textsc{DeltaPart3}$(I,c)$ computes $\Dfunapp{I}{c}$}
\label{alg:Delta}
\begin{algorithmic}[1]
\Function{DeltaPart3}{$I,c$} 
\Comment{Computes $\Dfunapp{I}{c}$}
%\If {$I = \emptyset$}  
%\State \Return $\lambda_{\top}(c)$
\If{$I =  \{ i \}$} 
\State\Return{$\sfunapp{i}{c}$} 
\Else \State$\ \{ J,K \} \gets \textsc{Partition}(I)$ 
\Comment{returns a partition $\{ J,K \}$ of $I$ s.t., $|J|=\lfloor{| I |/2} \rfloor$}
\State \Return $    \bigmeet\{   \textsc{DeltaPart3}(J,a) \join \textsc{DeltaPart3}(K,a \to c)  \ | \  a \in \Con \mbox{ and } a \cleq c \}. $
\EndIf
\EndFunction
\end{algorithmic}
\end{algorithm}

\emph{Algorithms.} Notice \textsc{DeltaPart3}$(I,c)$ computes $\Dfunapp{I}{c}$
using  Thm.\ref{thm:comp-algo}(3). By modifying Line 6 with the corresponding
meet operations, we obtain two variants of  \textsc{DeltaPart3} that use,
instead of Thm.\ref{thm:comp-algo}(3),  the Properties
Thm.\ref{thm:comp-algo}(1) and Thm.\ref{thm:comp-algo}(2). We call them 
\textsc{DeltaPart1} and  \textsc{DeltaPart2}. 

\emph{Worst-case time complexity}. We assume that binary distributive lattice 
operations $\meet$, $\join$, and $\to$ are computed in $O(1)$ time. We also
assume a fixed group $I$ of size $m=|I|$ and express the time complexity for
computing  $\Dfun{I}$ in terms of $n = | \Con \ |$, the size of the  set of
constraints. The above-mentioned algorithms compute the value 
$\Dfunapp{I}{c}$.  The worst-case time complexity for computing the function
$\Dfun{I}$ is in $O(mn^{1+
2\log_2m})$ using  \textsc{DeltaPart1}, and $O(m n^{1+\log_2m})$ using
\textsc{DeltaPart2} and \textsc{DeltaPart3} \cite{guzman:hal-02172415}.

%Apart from the above results, which are reported in full detail in \cite{guzman:hal-02172415},
%Wecompute  $\Dfun{I}$ in 
%$O(m n^{\log_2(3)})$ thus representing a noteworthy improvement over the previous algorithms. 
%We will also discuss a new tight upper bound on the number of 
%homomorphisms over any given lattice: $O( (n+1)^2 + n!L_n(-1))$ where $L_n(x)$ is the
%Laguerre polynomial of degree $n$ in $x$. 

\section{Conclusions and Related Work}
\label{sec:ds}

We have highlighted some results about scs as semantic structures for spatially-distributed systems exhibiting epistemic behaviour. Our work
in scs have been inspired by the seminal work on epistemic logic for knowledge and group knowledge in \cite{halpern1990knowledge,fagin1995reasoning,halpern2004reasoning}. 
Meaningful families of structures from logic and economics such as Kripke structures and Aumann structures have been shown to be instances of scs~\cite{knight:hal-00761116}.  
Furthermore scs have been used to give semantics to modal logics and process calculi \cite{knight:hal-00761116,guzman:hal-01257113,haar:hal-01256984}. 

 In \cite{knight:hal-00761116} we introduced a spatial and epistemic process
calculus, called sccp, for reasoning about spatial information and knowledge distributed among
the agents of a system. In this work scs were introduced as the domain-theoretical structures to
represent spatial and epistemic information. These structures are also used in the denotational and operational semantics
of sccp processes.   In \cite{knight:hal-00761116}  we also provided operational
and denotational techniques for reasoning about the potentially infinite behaviour of spatial and epistemic processes. 

In \cite{guzman:hal-01257113,haar:hal-01256984} we developed the theory of
spatial constraint systems (scs) with extrusion to specify information and processes
moving from a space to another.   In \cite{guzman:hal-01675010,guzman:hal-01328188} scs with extrusion are used to give a novel algebraic characterization of the notion of normality in modal logic and to derive right inverse/reverse operators for modal languages. These results were applied to derive new expressiveness results for bisimilarity and well-established modal languages such as 
Hennessy-Milner logic, and linear-time temporal logic. 

In \cite{guzman:hal-01257113,guzman:hal-01328188,guzman:hal-01675010,haar:hal-01256984}
scs are used to reason about beliefs, lies and other epistemic utterances but
also restricted to a finite number of agents and individual, rather than
group, behaviour of agents.  

In \cite{guzman:hal-02172415} we developed semantic foundations and provided algorithms for reasoning about
the distributed information of possibly infinite groups  in multi-agents systems. We plan to
develop similar techniques for reasoning about other group phenomena in multi-agent systems from social sciences and computer music such as group
polarization \cite{Esteban:94:Econometrica} and group improvisation
\cite{Rueda2004}.

We have recently learnt that the fundamental operations of dilation and erosion from digital images and mathematical morphology \cite{Serra:1983:IAM} are space and projection functions, respectively. 
Dilations are applied to figures. Intuitively, figures that are very lightly drawn get thick when dilated. We are currently studying potential applications of distributed spaces in
mathematical morphology: E.g., for computing the greatest dilation under a given set of dilations.  Similarly, we are also studying  scs interpretations of other fundamental operations from mathematical morphology such as opening and closing. 

We conclude with some applications of scs in the development of ccp tools and languages.  In \cite{haar:hal-01328189,haar:hal-01673529} we described D-SPACES, an implementation of
scs that  provides property-checking methods as well as an implementation of a specific type of constraint systems  (boolean
algebras). 
In \cite{ramirez:hal-01934953} we used rewriting logic for specifying and analyzing ccp processes combining spatial and real-time behavior. These processes can run processes in different computational spaces while subject to real-time requirements. The real-time rewriting logic semantics is fully executable in Maude with the help of rewriting modulo SMT: partial information (i.e., constraints) in the specification is represented by quantifier-free formulas on the shared variables of the system that are under the control of SMT decision procedures.  The approach is used to symbolically analyze existential real-time reachability properties of process calculi in the presence of spatial hierarchies for sharing information and knowledge.
We also developed \texttt{dspacenet}, a multi-agent spatial and reactive ccp language for
programming academic forums\footnote{You can try \texttt{dspacenet} at \url{http://www.dspacenet.com}.}. The fundamental structure of \texttt{dspacenet} is that of \emph{space}:  
A space may contain  spatial and reactive ccp programs or other spaces. The fundamental operation of \texttt{dspacenet} is that of \emph{program posting}:  In each time unit, agents (users) can post spatial reactive ccp programs in the spaces they are allowed to do so. Currently \texttt{dspacenet} is used at Univ. Javeriana Cali for teaching spatial reactive declarative programming.

\bibliographystyle{eptcs.bst}
\bibliography{refs}

\end{document}